# Crowdsourcing the Policy Cycle[1]

JOHN PRPIĆ, Beedie School of Business - Simon Fraser University
ARAZ TAEIHAGH[2], City Futures Research Centre - University of New South Wales
JAMES MELTON, College of Business Administration - Central Michigan University

1. **INTRODUCTION**

Crowdsourcing is beginning to be used for policymaking. The "wisdom of crowds" [Surowiecki 2005], and crowdsourcing [Brabham 2008], are seen as new avenues that can shape all kinds of policy, from transportation policy [Nash 2009] to urban planning [Seltzer and Mahmoudi 2013], to climate policy (http://climatecolab.org). In general, many have high expectations for positive outcomes with crowdsourcing, and based on both anecdotal and empirical evidence, some of these expectations seem justified [Majchrzak and Malhotra 2013]. Yet, to our knowledge, research has yet to emerge that unpacks the different forms of crowdsourcing in light of each stage of the well-established policy cycle. This work addresses this research gap, and in doing so brings increased nuance to the application of crowdsourcing techniques for policymaking.

2. **CROWDSOURCING COLLABORATION**

One perspective on collective intelligence views it as one of three possible types of IT-mediated crowdsourcing collaboration [de Vreede et al. 2009]. We employ and adapt the typology supplied by de Vreede et al. [2009], and detail the three types of crowdsourcing below.

**Virtual Labor Marketplaces**
A virtual labor marketplace (VLM) is an IT-mediated market for spot labor, where individuals and organizations can agree to execute work in exchange for monetary compensation. This type of crowdsourcing is typified by endeavors like Amazon's M-Turk and Crowdflower. The crowd of workers at these web properties are generally thought to excel at microtasks, such as the translation of documents, labelling photos, and participating in surveys [Narula et al. 2011], though they are not necessarily limited to such work. The crowd of laborers at these marketplaces are anonymous in respect to their offline identities and self-select the tasks that they are willing to undertake based upon the compensation offered for the task and the nature of the task itself. Given the size of the crowd at these marketplaces (for example, Crowdflower has over 5 million laborers – see http://crowdflower.com), tasks can be rapidly completed.

**Tournament-Based Collaboration**
In tournament-based collaboration (TBC) organizations post their problems or opportunities to IT-mediated crowds at web properties such as Innocentive and Kaggle [Afuah and Tucci 2012]. In posting a problem, the organization creates a competition amongst the crowd, where the best solution will be chosen as determined by the organization. The crowd of participants at these sites is generally smaller when compared to the VLM's (for example, Kaggle has approximately 140,000 data scientists that comprise its crowd -- see http://www.kaggle.com/solutions/connect), and the individual participants can choose not to be anonymous at these sites in relation to their offline identities. Fixed amounts of prize money are offered to the crowd for the winning solution and can range from a few hundred dollars to a million dollars or more (http://www.innocentive.com/files/node/casestudy/case-study-prize4life.pdf).

**Open Collaboration**
In an open collaboration (OC) model, organizations post their problem to the public at large through IT. Contributions from the crowds in these endeavors are voluntary and do not require monetary exchange. Posting on Reddit, starting a wiki, or using social media are examples of this type of collaboration. The scale of the crowds available to these types of endeavors can vary significantly depending on the reach and engagement of the IT used. For example, as of December 2013, Reddit had approximately 2.7 million registered Redditors (http://www.reddit.com/about) and though this mostly anonymous crowd is quite large, there is little to guarantee the attention of any significant subset of the contributors when using Reddit. Further, politicians such as Narendra Modi of India have in place very large personal communities of followers at Facebook (https://www.facebook.com/narendramodi) to the tune of 7.8 million "likes". On the other hand, California Assemblyman Mike Gatto has thus far garnered almost no response in his effort to crowdsource probate legislation (http://mikegatto.wikispaces.com). In short, unlike the other two methods of crowdsourcing already discussed here, the size of the crowds accessed in OCs can vary significantly.

---

[1] This work is the product of an exercise in Collective Intelligence creation undertaken at the HICSS 2014 workshop on "Crowdsourcing and Collective Intelligence" led by Jeff Nickerson. Big thanks to Jeff for leading a wonderful workshop, without which this work would not exist. Further thanks to KD Joshi for her participation in our deliberations, and to all the other workshop participants for sharing their expertise.

[2] Corresponding Author Email: a.taeihagh@unsw.edu.au, Address: City Futures Research Centre, FBE, UNSW, Sydney 2052, NSW, AUSTRALIA.

In Table #1 below, we compare the three types of crowdsourcing discussed here across three common characteristics. This set of characteristics reflects a minimum consensus extracted from the extant literature [de Vreede et al. 2009, Estellés-Arolas and Ladrón-de-Guevara 2012] and does not represent an exhaustive set of common characteristics. As far as we know, our use of this particular minimum set of common characteristics is the first of its kind, and we limit our analysis in this way, with the hope of providing a solid basis for further analysis.

In our comparison below, where possible, we use three-point estimates for each characteristic. In said comparison, cost refers to the nature of the direct expense involved for an implementer (individual or organization) to engage the crowd with each form of crowdsourcing. Potential auxiliary expenses, such as advertising and promotion of the crowdsourcing effort are not included in our estimates. Anonymity refers to the identity of the individuals in the crowd, as found within the IT used to engage the crowd and in relation to their offline identity. If there is a 1-to-1 correspondence of a crowd-member's online and offline identity, then anonymity would be low. Scale refers to the size of the crowd generally available to an implementer through each form of crowdsourcing collaboration.

Table #1 – Comparison of Types of Crowdsourcing

|  | Common Characteristics | Cost | Anonymity | Scale of Crowd |
|---|---|---|---|---|
| **Models of Crowdsourcing** |  |  |  |  |
| Virtual Labor-Markets |  | Variable | High | High |
| Tournament-Based Collaboration |  | Fixed | Medium | Medium |
| Open Collaboration |  | Free | Variable | Variable |

3. THE POLICY CYCLE

A policy is a set of effective and acceptable courses of action to reach explicit goals [Bridgman and Davis 2004]. The assumption is that policy makers are rational, though this assumption has been vigorously debated by some [Kingdon 1984, Stone 2002]. Systemic perspectives were first used for explaining political processes by Easton [1979], where political systems serve to convert inputs, such as political demands and public support, into outputs (i.e. a set of resulting decisions and actions). Palmer [1997] extended the application of systems to policies, conceiving the policy cycle as a relatively independent and interacting set of "blocks" having policy measures as inputs and a set of desired outcomes as outputs, which can be represented through causal diagrams. A policy cycle (see Figure #1) is defined as a sequence of steps in which an agenda is set; a problem is defined; alternative policies to address the problem are designed, analyzed and refined; a proposed policy is selected, implemented, enforced, and henceforth re-evaluated, challenged and/or revised [Stone 1988, Howlett et al. 1995].

**Figure #1 -** The Policy Cycle [In Taeihagh et al., 2009, adapted from Howlett et al. 1995]

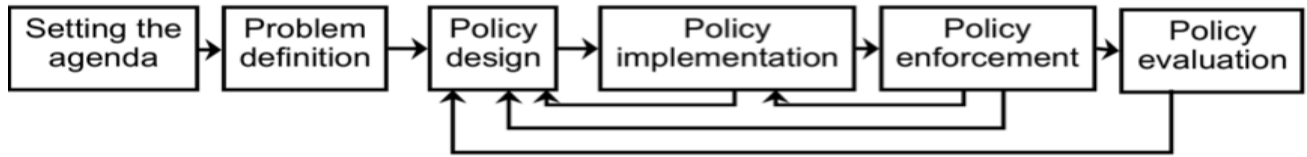

## 4. ANALYSIS and DISCUSSION

The different models of crowdsourcing outlined have different potentials and constraints as highlighted in Table #1. Similarly, the different blocks of the policy cycle have different input needs. Below, we examine each of the three modes of crowdsourcing and delineate the stages of the policy cycle for which they may be most useful (see Figure #2 below for a visual rendering).

**Figure #2 –** Crowdsourcing Methods in Relation to the Policy Cycle[3]

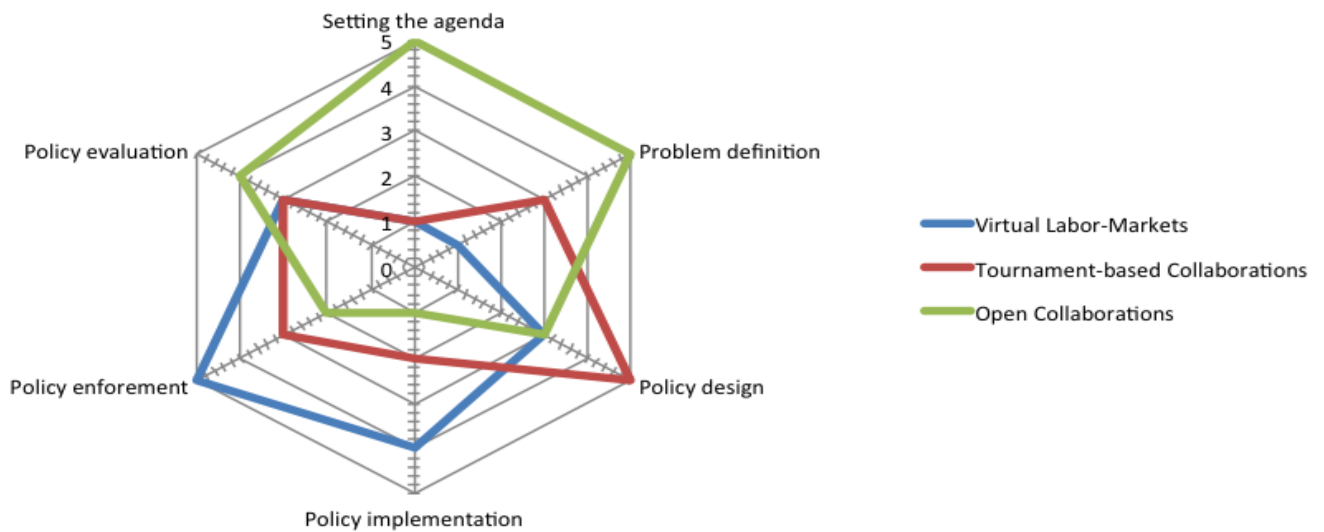

**Virtual Labor Marketplaces**
From our characterizations in Table #1 and in Figure #2, we can see that VLM's qualify (with scores of 3 or above in Figure#2) as potentially useful outlets for policy design, policy enforcement, policy implementation, and policy evaluation. For policy design, this stage of the policy cycle includes the generation of competing policy alternatives, and given the scale and attentiveness of the VLM crowds, it is likely that policy alternatives could be generated relatively quickly and cost-effectively. Further, if it happens that a particular policy design is in need of specialized skills or knowledge, workers can be filtered through a pre-qualification test at the VLM.

In terms of policy enforcement and implementation, given that the labor at VLM's can be employed to undertake offline tasks, they can be highly useful for policy enforcement. Applications such as Premise (http://www.premise.com) and FieldAgent (http://www.fieldagent.net) illustrate endeavors where crowds are used for offline tasks, and further illustrate that said tasks can be geographically segmented. In terms of policy evaluation, just as these VLM's are well-known as useful avenues for market research, it would stand to reason that policy evaluation (either prior or post-policy implementation) can similarly be actuated effectively, both in terms of cost and content.

---

[3] See Appendix #1 for a table listing the values that we assigned to create this depiction.

**Tournament-Based Collaboration**

From our characterizations in Table #1 and in Figure #2, we can see that TBCs qualify as useful outlets for policy design, policy enforcement, policy evaluation, and problem definition. For policy design, TBCs can be readily used to generate competing policy alternatives through contests at such web properties. Similarly such competitions could be used to ask the solvers at these crowds to generate new and useful metrics for policy evaluation. Further, given the highly specialized skills often found in these crowds, it may be that TBCs are a boon to the problem definition aspects of the policy cycle, where these crowds can step in after the policy agenda is set to assist in defining the problem very specifically or perhaps even in terms of formal mathematical models. In terms of policy enforcement, TBC's could be similarly set up, to generate new and innovative methods of enforcing policy.

**Open Collaboration**

From our characterizations in Table #1 and in Figure #2, we can see that OCs qualify as useful outlets for policy design, policy evaluation, problem definition, and agenda setting. If we assume a successful case of OC such as that illustrated by Narendra Modi of India, it appears that such a crowd can be a powerful tool for many aspects of the policy cycle. Given that each crowd member has voluntarily opted-in to following Mr. Modi and that they had to seek-out his profile to do so, the members of this crowd seem highly motivated to participate in his concerns, and may further share the very same geography that defines his constituency. In general, Mr. Modi could use his Facebook platform to canvass his "personal crowd" for all manner of issues, including the generation of policy alternatives and the evaluation of competing policy alternatives. In relation to the other types of crowdsourcing discussed here, it seems that Mr. Modi would have unique advantages in generating and setting agendas and problem definition, given the seemingly motivated and specialized nature of his personal crowd.

5. CONCLUSION & FUTURE WORK

It is important to recognize that the different phases of the policy cycle may benefit from the application of different types of crowdsourcing, and the work presented here is a tentative first step in this direction. In this work, we examined the three different forms of crowdsourcing, and examined them in light of the different stages of the policy cycle. From this initial analysis, it seems evident that crowdsourcing in general has the potential to have a positive impact in supporting and shaping the policy process.

In sum, given the illustrated differences among the types of crowdsourcing examined here, it appears that the different forms are suited to different roles and/or supporting functions in the policy process. Our Figure #2 highlights a speculative ranking of the different types of crowdsourcing for the different steps of the policy process. Given the complex nature of policy-making and the fact that policy issues are most often context-specific, we do not claim that our categorization and assessments are definitive. It's likely that there are many exceptions not captured by our analysis, including hybrid forms of crowdsourcing that cross-over our classification, as well as applications of the three crowdsourcing types that transcend the different stages of the policy cycle. This work provides the basis for further inquiry into these potential uses of crowdsourcing and a better understanding of the current limitations. Currently, we are conducting experiments in use of crowdsourcing for the policy process, and in general we suggest further inquiry into the following:

- A detailed characterization of the efficacy of the types of crowdsourcing based upon the policy cycle at the different levels of government (i.e. municipal, state or federal level);

- The application of crowdsourcing to different segments of the policy cycle;

- The application of multiple types of crowdsourcing on a single block of the policy cycle;

- An investigation into the preferences of different policy stakeholders for the use of different types of crowdsourcing in the policy process (E.G. A municipal government may want to use OC's for policy evaluation due to financial limitations, but a corporation looking to effect policy may prefer to utilize a TBC to better understand the impacts of a specific policy on their bottom-line).

# Appendix #1

**Table #2** - Suitability of the Types of Crowdsourcing for the Different steps of the Policy Cycle
(1= low, 5=high)

| Type | Setting the agenda | Problem definition | Policy design | Policy implementation | Policy enforcement | Policy evaluation |
|---|---|---|---|---|---|---|
| Virtual Labor-markets | 1 | 1 | 3 | 4 | 5 | 3 |
| Tournament-based collaborations | 1 | 3 | 5 | 2 | 3 | 3 |
| Open Collaborations | 5 | 5 | 3 | 1 | 2 | 4 |

REFERENCES

Allan Afuah and Christopher L. Tucci. 2012. Crowdsourcing as a solution to distant search. *Academy of Management Review.* 37, 3 (2012), 355-375.

Andrew Nash. 2009. Web 2.0 applications for improving public participation in transport planning, *Transportation Research Board 89th Annual Meeting*. (2009).

Anne Majchrzak and Arvind Malhotra. 2013. Towards an information systems perspective and research agenda on crowdsourcing for innovation. *The Journal of Strategic Information Systems*. 22, 4 (2013), 257-268.

Araz Taeihagh, Rene Bañares-Alcántara, and Claire Millican. 2009. Development of a novel framework for the design of transport policies to achieve environmental targets. *Computers & Chemical Engineering*. 33, 10 (2009), 1531-1545.

Bryan Palmer. 1997. Beyond program performance indicators: Performance information in a national system of health and family services. *Department of Health and Family Services*. Canberra, (1997).

Daren C. Brabham. 2008. Crowdsourcing as a model for problem solving: An introduction and cases. *Convergence: The International Journal of Research into New Media Technologies*. 14, 1, (2008), 75–90.

David Easton. 1979. *A systems analysis of political life.* Wiley, New York (1979).

Deborah Stone. 1988. *Policy paradox and political reason.* Harper Collins (1988)

Deborah Stone. 2002. *Policy paradox: The art of political decision making*. W. W. Norton & Company Ltd, New York (2002)

Enrique Estellés-Arolas and Fernando González-Ladrón-de-Guevara. 2012. Towards an integrated crowdsourcing definition. *Journal of Information Science.* 38, 2, (2012), 189-200.

Ethan Seltzer and Dillon Mahmoudi. 2013. Citizen Participation, Open Innovation, and Crowdsourcing Challenges and Opportunities for Planning. *Journal of Planning Literature*. 28, 1, (2013), 3-18.

Gert Jan de Vreede, Robert O. Briggs, and Anne P. Massey. 2009. Collaboration engineering: Foundations and opportunities. *Journal of the Association of Information Systems*, 10, 3, (2009), 121-137.

James Surowiecki. 2005. *The Wisdom of Crowds*. Anchor Books.

John W. Kingdon. 1984. *Agendas, Alternatives and Public Policy*. Little Brown, Boston, MA. (1984).

Michael Howlett, Michael Ramesh, and Anthony Perl. 1995. *Studying public policy: policy cycles and policy subsystems*. Oxford University Press, Toronto, (1995).

Peter Bridgman and Glyn Davis. 2004. *Australian Policy Handbook*. Allen & Unwin Academic, Sydney (2004).

Prayag Narula, Philipp Gutheim, David Rolnitzky, Anand Kulkarni, and Bjoern Hartmann. 2011. MobileWorks: A Mobile Crowdsourcing Platform for Workers at the Bottom of the Pyramid. *In Proceedings of HCOMP*. (2011).